\documentstyle[12pt,epsfig,graphics,cite,amssymb,amsmath,bm]{article}
\begin{document}\bibliographystyle{plain}\begin{titlepage}
\renewcommand{\thefootnote}{\fnsymbol{footnote}}\hfill
\begin{tabular}{l}HEPHY-PUB 968/16\\UWThPh-2016-12\\October
2016\end{tabular}\\[2cm]\Large\begin{center}{\bf INSTANTANEOUS
BETHE--SALPETER VIEW OF GOLDSTONE-TYPE PSEUDOSCALAR MESONS}\\[1cm]
\large{\bf Wolfgang LUCHA\footnote[1]{\normalsize\ {\em E-mail
address\/}: wolfgang.lucha@oeaw.ac.at}}\\[.3cm]\normalsize
Institute for High Energy Physics,\\Austrian Academy of
Sciences,\\Nikolsdorfergasse 18, A-1050 Vienna, Austria\\[1cm]
\large{\bf Franz F.~SCH\"OBERL\footnote[2]{\normalsize\ {\em
E-mail address\/}: franz.schoeberl@univie.ac.at}}\\[.3cm]
\normalsize Faculty of Physics, University of Vienna,\\
Boltzmanngasse 5, A-1090 Vienna, Austria\\[2cm]{\normalsize\bf
Abstract}\end{center}\normalsize

\noindent Describing the lightest pseudoscalar mesons as bound
states of quark and antiquark within the framework of an
instantaneous Bethe--Salpeter formalism constructed such as to
retain (in contrast to Salpeter's equation) as much information on
the relativistic effects provided by the full quark propagator as
conceivable allows for a surprisingly simple implementation of
their near masslessness mandatory for their interpretability as
pseudo-Goldstone bosons related to the spontaneous breaking of the
chiral symmetries of quantum chromodynamics.\vspace{3ex}

\renewcommand{\thefootnote}{\arabic{footnote}}\end{titlepage}

\section{Introduction}The pseudo-Goldstone bosons of the
spontaneously and to a minor degree explicitly broken chiral
symmetries of quantum chromodynamics (QCD), the quantum field
theory encoding all strong interactions in elementary particle
theory, are commonly identified with the light pseudoscalar
mesons, the pions and kaons. As quark--antiquark bound states,
these may be described, in principle, by the Bethe--Salpeter
formalism \cite{BSE}; in practice, such attempt~faces several
well-known difficulties. In view of this, we devised
\cite{WL05:LS} a three-dimensional reduction (the most famous of
this kind still is Salpeter's equation \cite{SE}) of the
Bethe--Salpeter equation by considering the instantaneous limit of
the latter but allowing for approximations only to the extent
definitely necessary for the analytic formulation of our
bound-state equation \cite{WL05:LS}.

In order to elucidate the relationship between the
pseudo-Goldstone bosons of QCD, on the one hand, and their
QCD-based description within the (instantaneous) Bethe--Salpeter
framework, on the other hand, we recently devised an adequate
inversion procedure \cite{WL13} and applied it \cite{WL13p} to the
maybe conceptually simplest three-dimensional reduction,
the~Salpeter equation, for which we analysed, in turn, the impact
of implementing the proper ultraviolet \emph{asymptotic
behaviour\/} \cite{WL15}, of constraints imposed by
\emph{axiomatic quantum field theory\/} \cite{WL16:ARP}, and of
exploiting available \emph{numerical knowledge\/} \cite{WL16:DSE}
of, in all three cases, the Goldstone~solution. Here, we
generalize these Goldstone-boson analyses to the bound-state
equation of Ref.~\cite{WL05:LS}.

The outline of this paper is as follows. In Sec.~\ref{Sec:IL}, we
briefly sketch the cornerstones of the instantaneous
Bethe--Salpeter formalism developed in Ref.~\cite{WL05:LS}, with
particular focus on the physical scenario relevant here. In
Sec.~\ref{Sec:GS}, we generalize the inversion technique of
Ref.~\cite{WL13} to the bound-state equation derived in
Ref.~\cite{WL05:LS}. In Sec.~\ref{Sec:PV}, aided by a
Ward--Takahashi~identity, we relate the Goldstonic Bethe--Salpeter
solution to the shape of the full quark propagator. In
Sec.~\ref{Sec:PR}, we derive the potential entering in our
instantaneous interaction kernel. In Sec.~\ref{Sec:SFCP}, we
summarize our findings. (For convenience, we use natural~units
throughout: $\hbar=c=1.$)

\section{Instantaneous Limit of Bethe--Salpeter Framework}
\label{Sec:IL}

\subsection{Dressed-propagator instantaneous Bethe--Salpeter
equation}Within the Poincar\'e-covariant Bethe--Salpeter formalism
\cite{BSE}, a bound state of two particles of momenta $p_{1,2}$ is
characterized by a momentum-dependent Bethe--Salpeter amplitude,
$\Phi.$ The basic idea behind our approach \cite{WL05:LS} is to
describe fermion--antifermion bound states, for total momentum $P$
and relative momentum $p$ of their constituents and mass
eigenvalues~$\widehat M$ defined by $P^2=\widehat M^2,$ by a
three-dimensional reduction of the homogeneous Bethe--Salpeter
equation which, nevertheless, retains relativistic effects to the
utmost possible extent. Such static reduction becomes possible if
assuming all underlying interactions to be independent of the time
components of the relative fermion momenta in the
center-of-momentum frame. The sole obstacles to this are then the
full fermion propagators entering in the bound-state equation. Any
fermion propagator $S(p)$ is defined by two Lorentz-scalar
functions~that may be chosen to be a mass function, $M(p^2),$ and
a wave-function renormalization factor,~$Z(p^2)$:
$$S_i(p)=\frac{{\rm i}\,Z_i(p^2)}{\not\!p-M_i(p^2)+{\rm
i}\,\varepsilon}\ ,\qquad\not\!p\equiv p^\mu\,\gamma_\mu\
,\qquad\varepsilon\downarrow0\ ,\qquad i=1,2\ .$$As a remedy, we
approximate the propagators $S_i(p)$ by retaining only terms
linear in $p_0$: an integration with respect to $p_0$ then gives a
bound-state equation \pagebreak for the Salpeter~amplitude
$$\phi(\bm{p})\equiv\frac{1}{2\pi}\int{\rm d}p_0\,\Phi(p,P)\ .$$

The considerations of Ref.~\cite{WL05:LS} resulted in the
instantaneous Bethe--Salpeter equation for fermion--antifermion
bound states with full propagators of the bound-state constituents
\cite{WL05:LS}
\begin{equation}\phi(\bm{p})=Z_1(\bm{p}_1^2)\,Z_2(\bm{p}_2^2)
\left(\frac{\Lambda_1^+(\bm{p}_1)\,\gamma_0\,I(\bm{p})\,\gamma_0\,
\Lambda_2^-(\bm{p}_2)}{P_0-E_1(\bm{p}_1)-E_2(\bm{p}_2)}-
\frac{\Lambda_1^-(\bm{p}_1)\,\gamma_0\,I(\bm{p})\,\gamma_0\,
\Lambda_2^+(\bm{p}_2)}{P_0+E_1(\bm{p}_1)+E_2(\bm{p}_2)}\right),
\label{Eq:3DR}\end{equation}wherein $E_i(\bm{p})$ and
$\Lambda_i^\pm(\bm{p})$ are the one-particle free energies and
energy projection operators
$$E_i(\bm{p})\equiv\sqrt{\bm{p}^2+M_i^2(\bm{p}^2)}\ ,\qquad
\Lambda_i^\pm(\bm{p})\equiv\frac{E_i(\bm{p})\pm
\gamma_0\,[\bm{\gamma}\cdot\bm{p}+M_i(\bm{p}^2)]}{2\,E_i(\bm{p})}
\ ,\qquad i=1,2\ ,$$and all interactions experienced by the
bound-state constituents are subsumed by the~term\begin{equation}
I(\bm{p})\equiv\frac{1}{(2\pi)^3}\int{\rm d}^3q\,K(\bm{p},\bm{q})
\,\phi(\bm{q})\ .\label{Eq:I}\end{equation}Of course, the
center-of-momentum frame of the two-particle system under
consideration is defined by $\bm{P}=0,$ whence $P_0=\widehat M$
and $\bm{p}=\bm{p}_1=-\bm{p}_2,$ and our bound-state
equation~reads
\begin{equation}\phi(\bm{p})=Z_1(\bm{p}^2)\,Z_2(\bm{p}^2)\left(
\frac{\Lambda_1^+(\bm{p})\,\gamma_0\,I(\bm{p})\,
\Lambda_2^-(\bm{p})\,\gamma_0}{\widehat M-E_1(\bm{p})-E_2(\bm{p})}
-\frac{\Lambda_1^-(\bm{p})\,\gamma_0\,I(\bm{p})\,
\Lambda_2^+(\bm{p})\,\gamma_0}{\widehat M+E_1(\bm{p})+E_2(\bm{p})}
\right).\label{Eq:CMF}\end{equation}

The particular projector structure of the right-hand side of our
bound-state equation, somewhat symbolically given by
$\Lambda_1^+\otimes\Lambda_2^-+\Lambda_1^-\otimes\Lambda_2^+,$
allows for a generally valid observation. Multiplying
Eqs.~(\ref{Eq:3DR}) or (\ref{Eq:CMF}) from both left and right by
appropriate energy projectors of the same sign cause their
right-hand sides to vanish. Hence, all solutions $\phi(\bm{p})$
inevitably~satisfy\begin{equation}\Lambda_1^+(\bm{p}_1)\,
\phi(\bm{p})\,\Lambda_2^+(\bm{p}_2)=\Lambda_1^-(\bm{p}_1)\,
\phi(\bm{p})\,\Lambda_2^-(\bm{p}_2)=0\ .\label{Eq:C}\end{equation}

The implications of retaining, in the three-dimensional reduction
of the Bethe--Salpeter equation, relativistic effects as far as
conceivable\footnote{Upon the assumption that all bound-state
constituents propagate freely with effective masses $m_i,$~that
is, in the free-propagator limit $M_i(\bm{p}^2)\to m_i$ and
$Z_i(\bm{p}^2)\to1,$ constituting a still further step of
simplifying approximation, the bound-state equation (\ref{Eq:CMF})
reduces to --- and, thus, generalizes --- Salpeter's equation
\cite{SE}.} by use of Eq.~(\ref{Eq:3DR}) for the
(semirelativistic) description \cite{WL92} of hadrons as bound
states of quarks \cite{WL91} have been analysed in
Refs.~\cite{WL05:DQ,WL06:C7}: comparison with earlier studies of
Salpeter's equation \cite{WL00a,WL00b,WL00c,WL01} revealed that
the inclusion of full quark propagators exerts a substantial
impact on the mass spectra of the bound~states.

\subsection{Spin-singlet fermion--antifermion states: Salpeter
amplitude}Bearing in mind the fact that the actual targets of the
present study, the pseudo-Goldstone bosons of QCD, are
\emph{pseudoscalar\/} mesons, let us recall\footnote{The
appropriate reasoning resembles the one used in, e.g.,
Refs.~\cite{LOVW} for the case of Salpeter's equation.} the
\emph{general\/} Lorentz structure of~the Salpeter amplitude for
bound states of fermion and antifermion whose spins add~up
to~zero.

The expansion of an arbitrary Salpeter amplitude over the basis
generated by the Dirac matrices introduces 16 component functions.
However, by the constraint (\ref{Eq:C}) this number is cut down by
a factor of two. Therefore, the most general solution
$\phi(\bm{p})$ of Eqs.~(\ref{Eq:3DR})~or~(\ref{Eq:CMF})~can
involve merely eight independent components. The latter differ in
their response to discrete symmetry transformations: Out of these
eight independent components, only two show the behaviour expected
for the spin-singlet bound states we are interested in.
Calling~these two components $\varphi_{1,2}(\bm{p})$ and imposing
flavour symmetry by requiring the fermion mass~equality
$M_1(\bm{p}^2)=M_2(\bm{p}^2)=M(\bm{p}^2),$ in the
center-of-momentum frame of our fermion--antifermion system under
consideration the general solution for spin-singlet states has to
be of the~form
\begin{equation}\phi(\bm{p})=\left[\varphi_1(\bm{p})\,
\frac{\gamma_0\,[\bm{\gamma}\cdot\bm{p}+M(\bm{p}^2)]}{E(\bm{p})}
+\varphi_2(\bm{p})\right]\gamma_5\ ,\qquad E(\bm{p})\equiv
\sqrt{\bm{p}^2+M^2(\bm{p}^2)}\ .\label{Eq:PSA}\end{equation}
Pseudoscalar states correspond to zero relative orbital angular
momentum of the fermions.

\subsection{Flavour-, Fierz- and spherical-symmetry-enabled
reductions}Now, in order to convert our three-dimensional
bound-state equation, given in Eq.~(\ref{Eq:CMF}) in its
momentum-space representation, to a well-defined manageable
inversion problem, we have to provide the particular
specifications of the physical scenario under consideration. To
this end, let us formulate a few simplifying assumptions emerging
from symmetry requirements that are dictated by physics or, at
least, may be justified on physical grounds. Our ultimate goal
will be to characterize the effective interactions responsible for
the formation of bound states in configuration-space
representation by some central potential~$V(r),$ where
$r\equiv|\bm{x}|.$\begin{enumerate}\item Flavour symmetry implies
equality both of all wave-function renormalization factors,
$$Z_1(\bm{p}^2)= Z_2(\bm{p}^2)=Z(\bm{p}^2)\ ,$$and of all mass
functions, with clear consequences for kinetic energies and
projectors:$$M_1(\bm{p}^2)=M_2(\bm{p}^2)=M(\bm{p}^2)\ ,\quad\!
E_1(\bm{p})=E_2(\bm{p})=E(\bm{p})\ ,\quad\!\Lambda_1^\pm(\bm{p})=
\Lambda_2^\pm(\bm{p})=\Lambda^\pm(\bm{p})\ .$$So, the
flavour-symmetric limit of our full-propagator bound-state
equation (\ref{Eq:CMF}) reads\begin{equation}\phi(\bm{p})=
Z^2(\bm{p}^2)\left(\frac{\Lambda^+(\bm{p})\,\gamma_0\,I(\bm{p})\,
\Lambda^-(\bm{p})\,\gamma_0}{\widehat M-2\,E(\bm{p})}-
\frac{\Lambda^-(\bm{p})\,\gamma_0\, I(\bm{p})\,\Lambda^+(\bm{p})\,
\gamma_0}{\widehat M+2\,E(\bm{p})}\right).\label{Eq:F}
\end{equation}\item The integral kernel $K(\bm{p},\bm{q})$ in the
interaction term (\ref{Eq:I}) subsumes both Lorentz nature --- by
generalized Dirac matrices $\Gamma_{1,2}$ --- and momentum
dependence --- by associated Lorentz-scalar potential functions
$V_\Gamma(\bm{p},\bm{q})$ --- of the effective interaction
experienced by the bound-state constituents; for
$\Gamma_1=\Gamma_2=\Gamma,$ the action of $K(\bm{p},\bm{q})$ on
$\phi(\bm{p})$~reads$$K(\bm{p},\bm{q})\,\phi(\bm{q})=\sum_\Gamma
V_\Gamma(\bm{p},\bm{q})\,\Gamma\,\phi(\bm{q})\, \Gamma\ .$$Fierz
symmetry is established by relying on the linear combination of
tensor products\begin{equation}\Gamma\otimes\Gamma=\frac{1}{2}\,
(\gamma_\mu\otimes\gamma^\mu+ \gamma_5\otimes\gamma_5-1\otimes1)\
.\label{Eq:D}\end{equation}\item The concurrence of the
convolution nature of the interaction term $I(\bm{p})$ in
Eq.~(\ref{Eq:3DR}) and the spherical symmetry of the
momentum-space potential function $V(\bm{p},\bm{q})$ related to
the Lorentz structure (\ref{Eq:D}), enforced by the requirement
$V(\bm{p},\bm{q})=V((\bm{p}-\bm{q})^2),$ implies that the Fourier
transform of $V(\bm{p},\bm{q})$ is a configuration-space central
potential, $V(r).$ Clearly, this nice feature allows us to discard
all dependence on the angular variables.\end{enumerate}Within the
setting specified thereby, the dressed-propagator instantaneous
Bethe--Salpeter equation (\ref{Eq:CMF}) may be easily shown to
reduce to the following system of coupled equations~for the radial
factors $\varphi_{1,2}(p),$ $p\equiv|\bm{p}|,$ of the independent
components $\varphi_{1,2}(\bm{p})$ required by $\phi(\bm{p}),$
\begin{subequations}\begin{align}&2\,E(p)\,\varphi_2(p)+
2\,Z^2(p^2)\int\limits_0^\infty \frac{{\rm d}q\,q^2}{(2\pi)^2}\,
V(p,q)\,\varphi_2(q)=\widehat M\,\varphi_1(p)\
,\label{Eq:5esa}\\&2\,E(p)\,\varphi_1(p)=\widehat M\,\varphi_2(p)\
,\label{Eq:5esb}\end{align}\end{subequations}in apposite notation:
$Z(\bm{p}^2)=Z(p^2),$ $M(\bm{p}^2)=M(p^2)$ and
$E(\bm{p})=E(p)\equiv\sqrt{p^2+M^2(p^2)}.$ These relations pose an
eigenvalue problem for the masses $\widehat M\equiv\sqrt{P^2}$ of
the bound~states in our focus of interest, wherein the
interactions of the constituents enter under the~disguise~of
$$V(p,q)\equiv\frac{8\pi}{p\,q}\int\limits_0^\infty{\rm d}r
\sin(p\,r)\sin(q\,r)\,V(r)\ ,\qquad q\equiv|\bm{q}|\ .$$The set of
relations formed by Eqs.~(\ref{Eq:5esa},\ref{Eq:5esb}) might be
considered as a kind of quintessence of our instantaneous
bound-state formalism for the particular physical system
defined~before.

\section{Exact Inversive Instantaneous Goldstone Solutions}
\label{Sec:GS}Next, in order to tighten the noose around the
precise way the strong interactions manifest, for the
pseudo-Goldstone bosons of QCD, in our instantaneous description
of bound states, let us explore the implications of the presumably
most eye-catching property of particles of this kind, namely, the
masslessness of Goldstone bosons requested by Goldstone's theorem.

Accordingly, we enumerate the not exorbitantly surprising
implications of enforcing, by letting $\widehat M=0,$ the
Goldstone nature of each spin-singlet bound state obeying
Eqs.~(\ref{Eq:5esa},\ref{Eq:5esb}):\begin{enumerate}\item For
vanishing bound-state mass $\widehat M,$ the two quintessence
equations (\ref{Eq:5esa},\ref{Eq:5esb}) decouple.\item Equation
(\ref{Eq:5esb}), being of algebraic nature, reduces to the demand
$E(p)\,\varphi_1(p)=0$ that coerces one Salpeter component to
vanish: $\varphi_1(\bm{p})=0.$ Thus, each Salpeter amplitude
(\ref{Eq:PSA}) for massless spin-singlet fermion--antifermion
bound states emerging as a solution to the flavour-symmetric limit
(\ref{Eq:F}) of the instantaneous bound-state equation
(\ref{Eq:3DR}) with Fierz-symmetric Dirac structure (\ref{Eq:D})
of its interaction kernel assumes the simple form$$\phi(\bm{p})=
\varphi_2(\bm{p})\,\gamma_5\ .$$\item Equation (\ref{Eq:5esa}), an
integral equation reminiscent of the relevant interactions,
becomes\begin{equation}E(p)\,\varphi_2(p)+Z^2(p^2)\int
\limits_0^\infty\frac{{\rm d}q\,q^2}{(2\pi)^2}\,V(p,q)\,
\varphi_2(q)=0\ ,\label{Eq:R}\end{equation}which controls the
surviving component $\varphi_2(\bm{p})$ and thus the Salpeter
amplitude $\phi(\bm{p}).$\end{enumerate}

Needless to say, the envisaged extraction of the interaction
potential $V(r)$ requires us to move, by means of a Fourier
transformation,\footnote{For functions of \emph{radial\/}
coordinates, Fourier transformations reduce to Fourier--Bessel
transformations.} to configuration space. Anticipating~that, for
finite momenta $p,$ the wave-function renormalization factor,
$Z(p^2),$ will prove to be~nonzero [$0\lneqq Z(p^2)\le1$], we
divide Eq.~(\ref{Eq:R}) by $Z^2(p^2).$ By introducing the
Fourier--Bessel transforms \pagebreak
\begin{align}\varphi(r)&\equiv\sqrt\frac{2}{\pi}\,\frac{1}{r}
\int\limits_0^\infty{\rm d}p\,p\sin(p\,r)\,\varphi_2(p) \
,\nonumber\\\widetilde T(r)&\equiv\sqrt\frac{2}{\pi}\,\frac{1}{r}
\int\limits_0^\infty{\rm d}p\,p\sin(p\,r)\,\frac{E(p)\,
\varphi_2(p)}{Z^2(p^2)}\label{Eq:FBT}\end{align}of the Salpeter
component $\varphi_2(p)$ and of the ratio of the kinetic term
$E(p)\,\varphi_2(p)$ and~$Z^2(p^2),$ we arrive at the sought
configuration-space representation of our bound-state
equation~(\ref{Eq:R}):$$\widetilde T(r)+V(r)\, \varphi(r)=0\
.$$From this, the Goldstone-promoting potential for the Lorentz
structure (\ref{Eq:D}) can be read~off:
\begin{equation}V(r)=-\frac{\widetilde T(r)}{\varphi(r)}\
.\label{Eq:Po}\end{equation}

\section{Quark-Propagator-Defined Bethe--Salpeter Vertex}
\label{Sec:PV}

From the aforegoing discussion, it should have become plain that
the envisaged application of the inversion procedure sketched
above requires as input the radial independent Salpeter component
$\varphi_2(p)$ that defines the Salpeter amplitude $\phi(\bm{p})$
of the bound states under study. By definition, the latter
quantity may be found from the Bethe--Salpeter amplitude
$\Phi(p,P)$ of these very bound states, by an integration with
respect to the time component $p_0$ of their constituents'
relative four-momentum $p.$ In the chiral limit, the
(renormalized) axial-vector Ward--Takahashi identity of QCD turns
out to relate the Bethe--Salpeter amplitude~$\Phi(p,0)$ for the
flavour-nonsinglet pseudoscalar mesons to the (dressed)
quark~propagator $S(p)$ \cite{PM97a}. This latter two-point Green
function, in turn, may be obtained as the solution to the quark
Dyson--Schwinger equation (occasionally also found to be referred
to as the gap~equation).\footnote{For good reasons,
Dyson--Schwinger analyses are preferentially performed in
Euclidean space; thus, we henceforth discriminate Euclidean-space
coordinates and Minkowski-space ones by underlining the~former.}

In Euclidean-space formulation, if skipping irrelevant overall
factors and focusing to the dominant Dirac component, the relation
between the Bethe--Salpeter amplitude
$\Phi(\underline{k},\underline{P})$~of a massless
($\underline{P}^2=-\widehat M^2=0$) pseudoscalar meson in its
center-of-momentum frame ($\underline{\bm{P}}=0$) and the
wave-function renormalization and mass functions
$Z(\underline{k}^2)$ and $M(\underline{k}^2)$ governing~the quark
propagator $S(\underline{k})$ has been proven (cf.~Sec.~IV.B of
Ref.~\cite{WL15}) to read, in the chiral~limit,
\begin{equation}\Phi(\underline{k},0)\propto
\frac{Z(\underline{k}^2)\,M(\underline{k}^2)}
{\underline{k}^2+M^2(\underline{k}^2)}\,\underline{\gamma}_5+
\mbox{subleading contributions}\ .\label{Eq:SP0}\end{equation}The
propagator functions are to be found by solving the quark
Dyson--Schwinger equation.

The Dyson--Schwinger equation for the quark propagator is an
element of the countably infinite hierarchy of the
Dyson--Schwinger equations of QCD, that is, of a system of coupled
integral equations that determines the infinity of $n$-point Green
functions of the~underlying quantum field theory. A manageable
Dyson--Schwinger problem may only be formulated by appropriately
decoupling, from this infinite tower, a finite subset of relations
which include the Dyson--Schwinger equation governing the Green
function of interest and continue to be coupled among each other.
Such a truncation cuts the links between subset and remainder:
necessary Green-function input has to be found by
phenomenologically inspired modelling.

In order to put us in a position to take advantage, at least at a
reasonable level~of~rigour, of the propagator--vertex
interrelation (\ref{Eq:SP0}) originating in the axial-vector
\pagebreak Ward--Takahashi identity of QCD, the symmetry
manifesting in form of this identity has to be respected also by
the finite number of Dyson--Schwinger equations singled out by the
favoured truncation. A class of models claimed to preserve the
identity is the rainbow--ladder truncation scheme likewise imposed
to quark Dyson--Schwinger equation and meson Bethe--Salpeter
equation. Its renormalization-group-improved formulation is
characterized by few generic properties:\begin{enumerate}\item The
exact \emph{quark--gluon vertex function\/} is substituted by its
tree-level approximation.\item The \emph{interaction kernel\/} is
cut down to its first perturbative term, one-gluon exchange.\item
The full \emph{gluon propagator\/} entering the gap equation is
approximated by its free~form.\end{enumerate}In order to
compensate, at least in part, for thereby induced deficiencies, an
effective~strong fine-structure coupling is introduced that takes
care of some phenomenological constraints:\begin{itemize}\item In
the infrared $\underline{k}^2\to0,$ it gets enhanced, to satisfy
the needs of the gluon propagator.\item In the ultraviolet
$\underline{k}^2\to\infty$, its decay parallels that of the
perturbative QCD coupling.\end{itemize}

\begin{figure}[hbt]\begin{center}\begin{tabular}{cc}
\psfig{figure=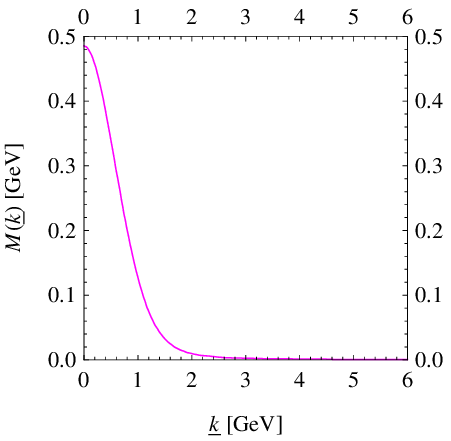,scale=1.680198}&
\psfig{figure=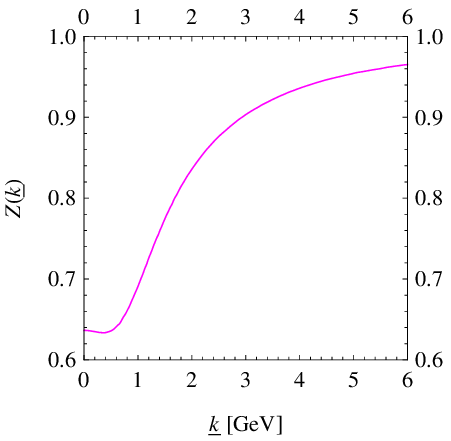,scale=1.680198}\\[1.171ex](a)&(b)\end{tabular}
\caption{Chiral-limit solutions to the Dyson--Schwinger equation
for the quark propagator $S(\underline{k})$ in the
``renormalization-group-improved'' rainbow--ladder model of
Ref.~\cite{M&T99}: (a) mass function $M(\underline{k})$ and (b)
wave-function renormalization $Z(\underline{k})$ read off from
Fig.~1 of~Ref.~\cite{PM00}.}\label{Fig:MZI}\end{center}
\end{figure}

For our present analysis, we rely on the model proposed in
Ref.~\cite{M&T99}, where the intended infrared enhancement of the
effective strong coupling strength is accomplished without the
introduction of a singular but integrable $\delta$ function in
momentum space. Unfortunately, for the moment the solution to the
Dyson--Schwinger equation for the quark propagator~can be computed
only in numerical form. Accordingly, we have to extract pointwise
the behaviour of both propagator functions $M(\underline{k}^2)$
and $Z(\underline{k}^2)$ from their graphs available in the
literature: Figure~\ref{Fig:MZI} shows the outcome of our simple
interpolation of the results published in Ref.~\cite{PM00}.

\section{Goldstone-Boson-Generating Interaction Potential}
\label{Sec:PR}Equipped, in form of Fig.~\ref{Fig:MZI}, with a
solution for the full quark propagator, we convert it into the
configuration-space potential $V(r).$ Being aware of the fact that
we have at our disposal only information on $M(\underline{k})$ and
$Z(\underline{k})$ from a limited interval of relative momenta
$\underline{k},$ we make every effort to underpin our intermediate
findings by pillars in form of analytic~expressions.

\begin{figure}[hbt]\begin{center}\begin{tabular}{cc}
\psfig{figure=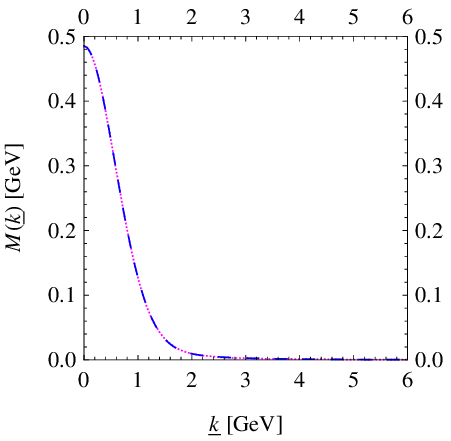,scale=1.680198}&
\psfig{figure=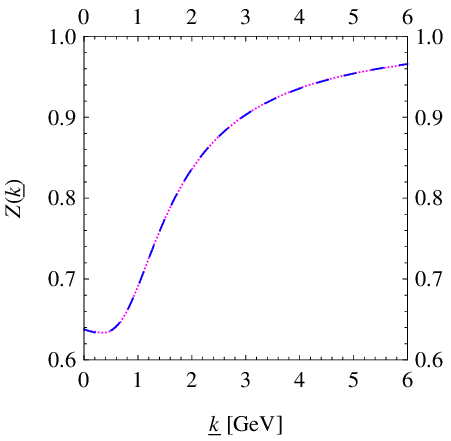,scale=1.680198}\\[1.171ex](a)&(b)\end{tabular}
\caption{Comparison of (a) the parametrization (\ref{Eq:M}) of the
mass function $M(\underline{k})$ for the five parameters of
Table~\ref{Tab:M} (blue dashed line) with its interpolation in
Fig.~\ref{Fig:MZI}(a) (magenta~dotted line) and (b) the
parametrization (\ref{Eq:Z}) of the wave-function renormalization
$Z(\underline{k})$ for the fit parameters of Table~\ref{Tab:Z}
(blue dashed line) with its origins in Fig.~\ref{Fig:MZI}(b)
(magenta dotted~line).}\label{Fig:MZP}\end{center}\end{figure}

First, we seek a parametrization of the propagator functions
$M(\underline{k})$ and $Z(\underline{k})$ in terms of elementary
functions. For the quark mass function, we rely on an ansatz
resembling~the~one of our Ref.~\cite{WL16:DSE}, with the numerical
values of the five parameters $a,b,\gamma,c,$ and $d$ as in
Table~\ref{Tab:M}:\begin{equation}M(\underline{k})
=\frac{a}{\left(1+\underline{k}^2/b\right)^\gamma}+c\,
\exp\!\left(-d\,\underline{k}^2\right).\label{Eq:M}\end{equation}
For the quark wave-function renormalization, we try a shape
complying with the behaviour $Z(\underline{k})\to1$ for
$\underline{k}\to\infty,$ finding for the six parameters
$u,v,x,y,z,$ and $\delta$ the values in Table~\ref{Tab:Z}:
\begin{equation}Z(\underline{k})=u-\frac{v}
{\left(1-x\,\underline{k}+y\,\underline{k}^2+z\,\underline{k}^4
\right)^\delta}\ .\label{Eq:Z}\end{equation}Figure~\ref{Fig:MZP}
confronts our parametrization of $M(\underline{k})$ and
$Z(\underline{k})$ by the ans\"atze (\ref{Eq:M}) and (\ref{Eq:Z})
with their interpolations given in Fig.~\ref{Fig:MZI}. Beyond
doubt, the agreement is more than~satisfactory.

\begin{table}[hb]\caption{Numerical values of the parameters in our
ansatz (\ref{Eq:M}) for the quark mass function.}\label{Tab:M}
\begin{center}\begin{tabular}{lrrrrr}\hline\hline\\[-1.5ex]
Parameter&\multicolumn{1}{c}{$a\left[\mbox{GeV}\right]$}
&\multicolumn{1}{c}{$b\left[\mbox{GeV}^2\right]$}
&\multicolumn{1}{c}{$\gamma$}
&\multicolumn{1}{c}{$c\left[\mbox{GeV}\right]$}
&\multicolumn{1}{c}{$d\left[\mbox{GeV}^{-2}\right]$}\\[1.5ex]\hline
\\[-1.5ex]Value&0.112918&0.870237&1.53153&0.371719&1.39261
\\[1.5ex]\hline\hline\end{tabular}\end{center}\end{table}

\begin{table}[b]\caption{Numerical parameter values fixing our
wave-function renormalization ansatz (\ref{Eq:Z}).}\label{Tab:Z}
\begin{center}\begin{tabular}{lrrrrrr}\hline\hline\\[-1.5ex]
Parameter&\multicolumn{1}{c}{$u$}&\multicolumn{1}{c}{$v$}
&\multicolumn{1}{c}{$x\left[\mbox{GeV}^{-1}\right]\!$}
&\multicolumn{1}{c}{$y\left[\mbox{GeV}^{-2}\right]\!$}
&\multicolumn{1}{c}{$z\left[\mbox{GeV}^{-4}\right]\!$}
&\multicolumn{1}{c}{$\delta$}\\[1.5ex]\hline
\\[-1.5ex]Value&1.01234&0.374334&0.229377&0.258882&0.592377&0.313777
\\[1.5ex]\hline\hline\end{tabular}\end{center}\end{table}

\begin{figure}[hbt]\begin{center}\begin{tabular}{cc}
\psfig{figure=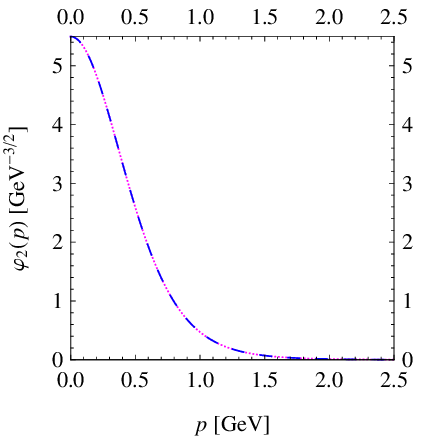,scale=1.72052}&
\psfig{figure=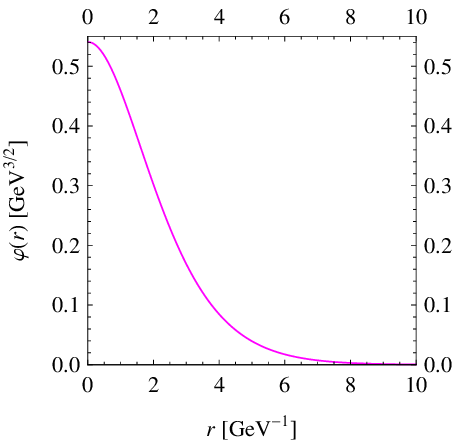,scale=1.72052}\\[1.171ex](a)&(b)\end{tabular}
\caption{ Radial Salpeter amplitude for Goldstone-type
quark--antiquark bound states: (a) its momentum-space behaviour
$\varphi_2(p),$ given by the outcome of the numerical integration
of its Bethe--Salpeter amplitude (\ref{Eq:SP0}) with respect to
Euclidean time (magenta dotted line)~or, equally well, the simple
parametrization (\ref{Eq:phip}) (blue dashed line); (b) its
configuration-space behaviour $\varphi(r),$ found as
Fourier--Bessel transform of that ansatz (\ref{Eq:phip}) (magenta
solid line).}\label{Fig:phi}\end{center}\end{figure}

Next, by exploiting the Dyson--Schwinger--Bethe--Salpeter
interplay (\ref{Eq:SP0}), we determine the center-of-momentum
Bethe--Salpeter amplitude (\ref{Eq:SP0}) and from this, by
integration with respect to the Euclidean-time coordinate
$\underline{k}_4,$ the sought radial component $\varphi_2(p),$
$p\equiv|\bm{p}|,$ of the three-dimensional Salpeter amplitude
$\phi(\bm{p}).$ The interpolated outcome of the numerical
integration is compared, in Fig.~\ref{Fig:phi}(a), with the simple
parametrization,~meaningful if $\eta>\frac{3}{4},$\begin{equation}
\varphi_2(p)=\sqrt{\frac{\Gamma(2\,\eta)}
{\sqrt{\pi}\,\Gamma(2\,\eta-\frac{3}{2})}}\,
\frac{2\,s^{2\,\eta-\frac{3}{2}}}{\left(p^2+s^2\right)^\eta}\
,\qquad\|\varphi_2\|^2\equiv\int\limits_0^\infty{\rm
d}p\,p^2\,|\varphi_2(p)|^2 =1\ ,\label{Eq:phip}\end{equation}for
the numerical values of the two parameters, $s$ and $\eta,$ in
Table~\ref{Tab:phip}. In spite of the~simplicity of this
parametrization, it exhibits almost perfect agreement with the
integration outcome.

\clearpage

\begin{table}[ht]\caption{Constants enabling us to reproduce the
Salpeter function $\varphi_2(p)$ by our ansatz~(\ref{Eq:phip}).}
\label{Tab:phip}\begin{center}\begin{tabular}{lrr}\hline\hline\\
[-1.5ex]Parameter&\multicolumn{1}{c}{$s\left[\mbox{GeV}\right]$}
&\multicolumn{1}{c}{$\eta$}\\[1.5ex]\hline\\[-1.5ex]Value&1.16176
&4.43353\\[1.5ex]\hline\hline\end{tabular}\end{center}\end{table}

\begin{figure}[hbt]\begin{center}\psfig{figure=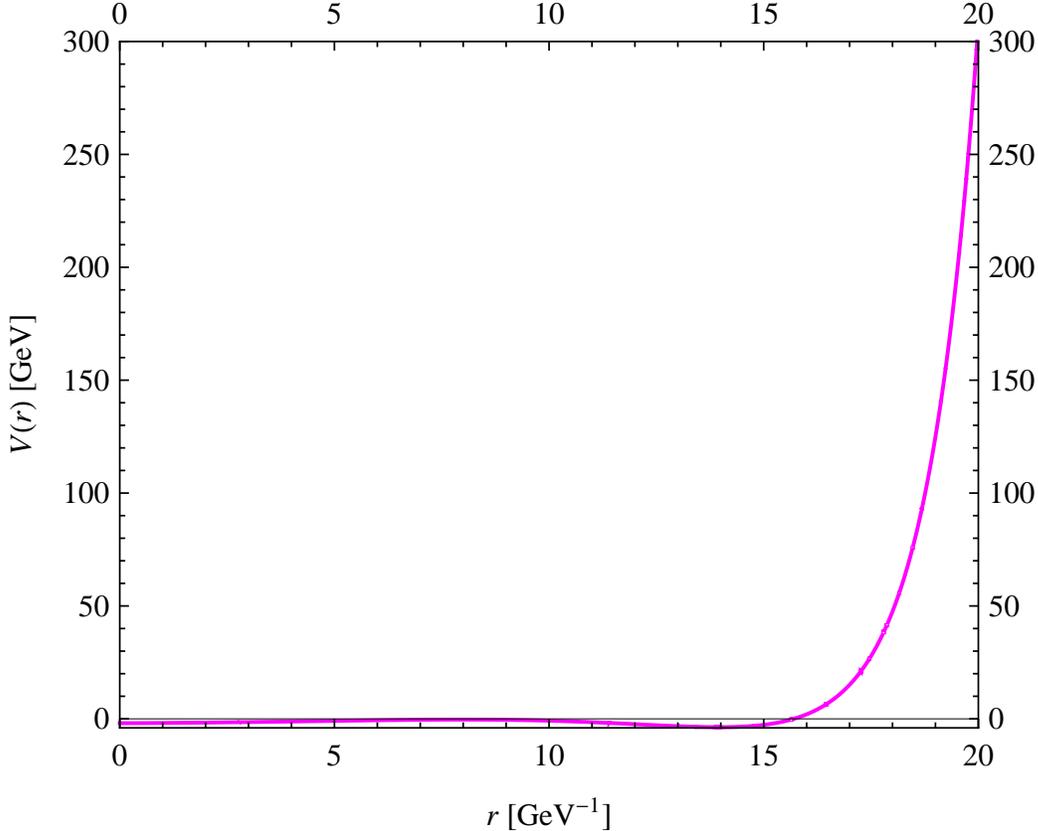,scale=1.8}
\caption{Configuration-space potential $V(r)$ derived by inverting
the bound-state problem posed by the
(``Salpeter-approach-improving'') instantaneous Bethe--Salpeter
equation (\ref{Eq:3DR}) for fermion--antifermion bound states
proposed in Ref.~\cite{WL05:LS}, with the Fierz-invariant Lorentz
structure (\ref{Eq:D}) of its interaction kernel, upon application
of our (Ward--Takahashi-mediated) knowledge of its solutions
describing nearly massless Goldstone-type pseudoscalar mesons.}
\label{Fig:V}\end{center}\end{figure}

As our final move, we clearly have to switch over, by
straightforward application of both Fourier--Bessel
transformations (\ref{Eq:FBT}), to configuration space. For the
function $\varphi_2(p),$ thanks to the simplicity of our
representation (\ref{Eq:phip}), this can be accomplished by
analytic~means. In terms of the modified Bessel functions of the
second kind of order $\sigma\in{\mathbb R}$ \cite{AS},
$K_\sigma(z),$~we~get$$\varphi(r)=\sqrt{\frac{\Gamma(2\,\eta)}
{\sqrt{\pi}\,\Gamma(2\,\eta-\frac{3}{2})}}\,
\frac{2^{2-\eta}\,s^\eta}{\Gamma(\eta)}\,
r^{\eta-\frac{3}{2}}\,K_{\frac{3}{2}-\eta}(s\,r)\ ,\qquad
\|\varphi\|^2\equiv\int\limits_0^\infty{\rm
d}r\,r^2\,|\varphi(r)|^2=1\ .$$The behaviour of $\varphi(r),$
depicted in Fig.~\ref{Fig:phi}(b), does not offer any unexpected
surprise. As far as the kinetic term is concerned, because of the
presence of the quark propagator functions, $M(p^2)$ and $Z(p^2),$
the Fourier--Bessel transform $T(r)$ has to be computed
numerically. This implies that for the underlying interaction
potential $V(r)$ no analytic representation~can be given.
Figure~\ref{Fig:V} shows its dependence on the interquark distance
$r$ as inferred from~Eq.~(\ref{Eq:Po}).

As a function of the interquark distance $r,$ the
configuration-space interaction potential $V(r)$ displayed in
Fig.~\ref{Fig:V} exhibits a not extraordinarily spectacular
behaviour: at the spatial origin $r=0,$ it starts from its
negative value $V(0)=-1.91772\;\mbox{GeV},$ then remains, without
very much variation or modulation, below zero until it crosses the
abscissa at its~single zero at $r=15.6978\;\mbox{GeV}^{-1},$ in
order to move on to an extremely steep monotonic rise to infinity,
bearing a coarse resemblance to a smoothed infinite square well,
and entailing~confinement.

\section{Summary, Findings, Conclusions, and Perspectives}
\label{Sec:SFCP}The present investigation was devoted to the
application of (the instantaneous limit of) the homogeneous
Bethe--Salpeter formalism to the description of the
pseudo-Goldstone bosons of the chiral symmetries of QCD, in an
insignificantly idealized disguise, namely, as strictly massless
mesons. In particular, we explored the respective capabilities of
the instantaneous Bethe--Salpeter equation formulated in
Ref.~\cite{WL05:LS} that (in spite of the reduction it underwent)
still has memories of sufficient intensity to its
quantum-field-theoretic origins to encompass effects decisive for
the understanding of the dynamical breakdown of the chiral
symmetries.

We were able to show that this specific bound-state equation
enables us to deal with the Goldstone bosons in a surprisingly
simple manner: As illustrated by Fig.~\ref{Fig:V}, the shape of
the configuration-space potential introducing the strong
interactions into the integral~kernel of this equation of motion
proves to be pretty close to a (bag-model-type) infinite square
well. In contrast to that, in our previous analyses of similar
applications of the Salpeter equation we encountered
parameter-dependent occurrences of singularities or nonconfinement
\cite{WL15,WL16:ARP,WL16:DSE}.

Moreover, in these earlier works we ignored the implications of
the quark wave-function renormalization factors, either because we
merely exploited one or the other specific aspect of the quark
mass functions \cite{WL15,WL16:ARP} or simply because of the lack
of availability of appropriate results in the literature for the
wave-function renormalization function from the underlying
Dyson--Schwinger truncation model \cite{WL16:DSE}. Here, the quark
wave-function renormalization has been taken into account, which
entails that the relationship between quark propagator~and
pseudoscalar-meson Bethe--Salpeter amplitude has been used more
accurately than before. A closer inspection reveals that this move
has discernible quantitative consequences for~the behaviour of the
potential but will hardly imply a dramatic qualitative impact on
its shape.

Of these insights clearly most essential is the square-well shape
of the potential.\footnote{Needless to say, in this context we use
the notion ``square well'' just as a metaphor for the $r$
dependence of the interquark potential $V(r)$ observed in
Sec.~\ref{Sec:PR}, namely, a rather flat behaviour near the origin
$r=0$ up to a critical distance beyond which, for $r\to\infty,$ an
extremely steep rise, with finite slope, to infinity~follows.}

Finally, a word of caution: By a too cursory interpretation of the
inversion outcomes for the interquark potential $V(r)$ shown in
Fig.~\ref{Fig:V}, one might na\"ively suspect that this potential
should describe mesonic bound states that are much larger than
what one would expect for the size of, e.g., the pion. However,
already by construction the ground-state solution of the
Bethe--Salpeter quintessence (\ref{Eq:R}) with the interquark
potential $V(r)$ of Fig.~\ref{Fig:V} is the~Salpeter component
wave function $\varphi(r)$ that provides the starting point of our
inversion adventures. (The latter claim can be easily verified by,
e.g., application of some variational techniques.) Realizing this,
from a brief inspection of the behaviour of $\varphi(r)$ given in
Fig.~\ref{Fig:phi}(b), one learns that the corresponding hadronic
extension should be of the order of some $2.5\;\mbox{GeV}^{-1}.$
More precisely, our $\varphi(r)$ --- and thus the bound-state
equation (\ref{Eq:R}) with interquark potential $V(r)$ as in
Fig.~\ref{Fig:V} --- describes massless mesons of average
interquark distance $\langle r\rangle=0.483\;\mbox{fm}$ and
root-mean-square radius $\sqrt{\langle r^2\rangle}=0.535\;
\mbox{fm}.$ Confronting our results with the experimental outcomes
\cite{PDG} for the electromagnetic charge radius of the pion
$\sqrt{\langle r_\pi^2\rangle}=(0.672\pm0.008)\;\mbox{fm},$ we
find a reasonable proximity of the numerical findings, lending
credibility to our analysis.

\end{document}